\documentstyle[10pt,emulateapj]{article}

\received{September 10 2000}
\accepted{January 18 2001}

\begin{document}

\title{Expectations for SZ cluster counts: mass function 
versus X-ray luminosity function}

\author{Yan-Jie Xue and Xiang-Ping Wu}

\affil{Beijing Astronomical Observatory and 
National Astronomical Observatories,
Chinese Academy of Sciences, Beijing 100012; China}

\begin{abstract}
We present a comparison of the SZ cluster counts predicted by 
the Press-Schechter (PS) mass function (MF) and the X-ray luminosity 
function (XLF) of clusters. 
The employment of the cluster XLF, together with the observationally 
determined X-ray luminosity($L_{\rm X}$)-temperature($T$) relation, 
may allow us to estimate the SZ cluster counts in a more realistic manner, 
although such an empirical approach depends sensitively on our
current knowledge of the dynamical properties of intracluster gas and its
cosmic evolution. Using both the non-evolving and evolving XLFs of clusters 
suggested by X-ray observations, we calculate the expectations for  
SZ surveys of clusters with X-ray luminosity  
$L_{\rm X}\geq 3\times10^{44}$ erg s$^{-1}$ and 
$L_{\rm X}\geq 1\times10^{43}$ erg s$^{-1}$ in
the 0.5 - 2.0 band, respectively. The non-evolving XLF results 
in a significant excess of SZ cluster counts at high redshifts as 
compared with the evolving XLF, while a slightly steeper 
$L_{\rm X}$-$T$ relation than the observed one is needed to
reproduce the distributions of SZ clusters predicted 
by the standard PS formalism. It is pointed out that uncertainties 
in the cosmological application of future SZ cluster surveys 
via the standard PS formalism should be carefully studied.
\end{abstract}


\section{Introduction}

There is growing evidence from X-ray observations and numerical 
simulations that bulk of the baryons in the universe resides 
in groups and clusters of galaxies, seen as diffuse ionized gas
(e.g. Fukugita, Hogan \& Peebles 1998). 
The hot gas interacts the passing cosmic microwave background (CMB) 
photons through the so-called Sunyaev-Zel'dovich (SZ) effect,
depending uniquely on the total thermal energy of 
the gas contained in the system. Consequently, the effect
is not only independent of the redshift 
of the system but also insensitive to detailed structure of 
the gas distribution. Therefore, it is well suited for studies of 
distant groups and clusters and their cosmic evolution. 
Motivated by these advantages, the purely SZ-based sky surveys
have been proposed (Bartlett \& Silk 1994; Barbosa et al. 1996)
and several dedicated interferometric arrays for non-targeted SZ surveys 
are under construction (e.g. Kneissl 2000; 
Holder et al. 2000;  Udomprasert, Mason \& Readhead 2000;
Fan \& Chiueh 2001).

It has been realized that non-targeted SZ surveys can act as a
sensitive probe of the cosmological parameters via the SZ counts
and in particular their redshift distribution (Barbosa et al. 1996).
Both numerical simulations (e.g. Holder et al. 2000;
Seljak, Burwell \& Pen 2000; Springel, White \&
Hernquist 2000) and analytic treatment (e.g. Bartlett 2000; 
Fan \& Chiueh 2001) have been developed in recent years towards the 
demonstration of how remarkably one can set tight constraints on the 
cosmic density parameter $\Omega_{\rm M}$, the cosmological constant
$\Omega_{\Lambda}$ and the normalization parameter $\sigma_8$ using the
forthcoming SZ surveys. Essentially, all these predictions are based
on the  Press-Schechter (PS) mass function (MF) 
(Press \& Schechter 1974), together with the mass-temperature relation and  
a constant gas fraction which convert the gas temperature and mass 
in the SZ effect into the dark halo mass in the PS formalism.
While  the PS formalism and the mass-temperature relation
have been justified by numerous optical/X-ray observations 
and in particular by 
numerical simulations on broad mass scales from groups to rich 
clusters, the robustness of our predictions for SZ source counts 
based purely on these theoretically motivated mechanisms should be
further tested. Recall that SZ surveys provide no direct 
information about the distribution and evolution of 
dark halos in the universe. Whenever the gas 
is involved, we should worry about the contamination of 
other non-gravitational heating processes, 
the possible departure of the gas from hydrostatic equilibrium, and
the different evolution scenario of the gas from that of the dark halos. 
Therefore, it is worthy of exploring an independent estimate of 
SZ counts based more closely on the well-established observational facts,
and comparing it with the result from the above standard theoretical model.
This permits an examination of the reliability/uncertainty of 
the theoretically expected SZ counts and their cosmological applications.

In this paper, we calculate the expectations for SZ surveys of clusters 
using the observationally determined cluster X-ray luminosity 
function (XLF) and X-ray luminosity($L_{\rm X}$)-temperature($T$) 
relation instead of the standard PS formalism. Our strategy is to utilize
as far as possible our current knowledge of X-ray clusters 
in the prediction of SZ cluster counts.
We wish to compare our prediction on the basis of observational facts with
that from the theoretical PS formalism. 
In fact, there is noticeable difference in the working hypotheses
between the two scenarios: First, the cluster XLF exhibits no or  
only mild evolution at least out to $z\approx1$, while the MF of dark  
halos represented by the PS formalism is sensitive to
redshift unless an open cosmological model or 
a nonzero cosmological constant is properly introduced; 
Second, the observed $L_{\rm X}$-$T$ relation for clusters 
($L_{\rm X}\propto T^3$)
deviates remarkably from the simply scaling law ($L_{\rm X}\propto T^2$) 
expected from self-similar evolution of both dark matter and intracluster 
gas with cosmic epoch,  which compares with the well established 
mass-temperature relation ($M\propto T^{3/2}$) 
adopted in the standard PS approach.  
In a sense, the employment of the observed $L_{\rm X}$-$T$ relation permits 
the inclusion of other non-gravitational heating  in the prediction
of SZ cluster counts, although the effect may be minor for rich clusters.
Alternatively, we note that the X-ray cluster counts can be 
reproduced remarkably well within the framework of the PS formalism
via the cluster $L_{\rm X}$-$T$ relation (Kitayama \& Suto 1997).  
It remains interesting to examine the issue of 
whether the PS formalism and the cluster XLF via the $L_{\rm X}$-$T$ 
relation can predict the same SZ cluster counts.

\section{SZ flux}

We begin with a brief summary of the procedure of evaluating the total SZ 
flux for a given cluster at redshift $z$ (see also Barbosa et al. 1996).
The change in CMB intensity by thermal SZ effect of hot intracluster
gas is
\begin{equation}
\Delta I_{\nu} = j_{\nu}(x)\;
        \int \sigma_{\rm T} n_{\rm e}\left(\frac{kT}{m_{\rm e}c^2}\right)d\ell,
\end{equation}
where $\sigma_{\rm T}$ is the Thomson cross-section, $m_{\rm e}$, 
$n_{\rm e}$ and $T$ are
the electron mass, density and temperature, respectively, and the integral
is along the line-of-sight. The spectral dependence is represented by
\begin{eqnarray}
j_{\nu}(x)=2\frac{(kT_{\rm CMB})^3}{(h_{\rm p}c)^2}f_{\nu}(x);\\
f_{\nu}(x)=\frac{x^4e^x}{(e^x-1)^2}
	   \left[x\coth\frac{x}{2}-4\right],
\end{eqnarray}
in which $x=h_{\rm p}\nu/kT_{\rm CMB}$, and $T_{\rm CMB}=2.728$ K  
is the temperature of CMB.
The total integrated SZ flux over the cluster face is then
\begin{equation}
S_{\nu}=\frac{j_{\nu}(x)}{D_a^2(z)}
        \left(\frac{\sigma_{\rm T}}{m_{\rm e}c^2}\right)
	\int kT n_{\rm e} dV,
\end{equation}
where $D_a(z)$ is the angular diameter distance to the cluster. 
If we assume that the gas is isothermal, then the above integral
over the entire volume of a cluster can be replaced by the
total gas mass of the cluster, $M_{\rm gas}$. Introducing the gas
fraction $f_b=M_{\rm gas}/M$, where $M$ is the total cluster mass,
we have 
\begin{equation}
S_{\nu}=\frac{j_{\nu}(x)}{D_a^2(z)}
        \left(\frac{kT}{m_{\rm e}c^2}\right)
	\left(\frac{f_b\sigma_{\rm T}}{\mu_{\rm e}m_{\rm p}}\right) \; M.
\end{equation}

Virial theorem suggests that the virial mass $M$, virial radius $r_{\rm vir}$
and temperature $T$ of a cluster are scaled as $kT\propto GM/r_{\rm vir}$.
Specifically, if we define $r_{\rm vir}$ as the radius within which the mean
cluster density is $\Delta_c$ times the critical density of the universe
$\rho_c$, i.e., 
$M=4\pi r_{\rm vir}^3 \Delta_c\rho_c/3$, then the mass-temperature
relation in terms of the notations of Bryan \& Norman (1998) reads
\begin{eqnarray}
kT=kT^*\left(\frac{M}{10^{15}M_{\odot}}\right)^{2/3};\\
kT^*=1.39f_{\rm T}(h^2\Delta_cE^2)^{1/3} \;{\rm keV},
\end{eqnarray}
where $H(z)=100hE(z)$ is the Hubble constant,  
$E^2(z)=\Omega_{\rm M}(1+z)^3+\Omega_{\rm K}(1+z)^2+\Omega_{\Lambda}$, 
$\Omega_{\rm K}=1-\Omega_{\rm M}-\Omega_{\Lambda}$, and 
$f_{\rm T}$ is a normalization factor in the range of
$0.75<f_{\rm T}<0.92$.  
This relation has been well justified by both numerical simulations 
(Evrard, Metzler \& Navarro 1996; Bryan \& Norman 1998) and
X-ray observations (Horner et al. 1999).

The X-ray luminosity $L_{\rm X}$ emitted 
as thermal bremsstrahlung by the intracluster gas follows
$L_{\rm X}\propto n_{\rm e}^2T^{1/2}r_{\rm vir}^3$. In the self-similar model
(Kaiser 1986), this reduces to $L_{\rm X}\propto f_b^2 T^2$.
In principle, the X-ray luminosity $L_{\rm X}$ can be linked with
the virial mass $M$ through equation (6). 
However, there are primarily two reasons that we cannot adopt this
approach in our analysis below: Firstly, unlike the $M$-$T$ relation
discussed above, it remains unclear  how to properly normalize
the $L_{\rm X}$-$T$ relation (see Kitayama \& Suto 1997). 
Secondly, 
a number of observational tests of the $L_{\rm X}$-$T$ relation 
have revealed a steeper $L_{\rm X}$-$T$ relation 
close to $L_{\rm X}\propto T^{3}$ especially in poor clusters
(see Wu, Xue \& Fang 1999 for a recent summary), which deviates
significantly from the theoretical argument $L_{\rm X}\propto T^2$, 
provided that the baryon fraction in clusters is universal. 
Such a discrepancy is often interpreted as the contamination of 
non-gravitational heating processes in the early phase of formation
and evolution of clusters  (e.g. Cavaliere, Menci \& Tozzi 1998; 
Wu, Fabian \& Nulsen 1998; Ponman, Cannon \& Navarro 1999).
If we confine ourselves to rich
clusters as selected by the ongoing and upcoming SZ surveys, 
the effect of non-gravitational heating is probably only minor.
Nonetheless, without loss of generality we adopt the following 
$L_{\rm X}$-$T$ relation determined empirically from X-ray observations
\begin{equation}
kT=kT_0 \left(\frac{L_{\rm X}}{L_0}\right)^B (1+z)^C,
\end{equation}
where the free parameters $T_0$, $L_0$, $B$  and $C$ can be fixed
in the fitting of X-ray data.

Now, the total SZ flux equation (5) can be re-written as
\begin{equation}
S_{\nu}=S_0\left(\frac{M}{10^{15}M_{\odot}}\right)^{5/3},
\end{equation}
or
\begin{equation}
S_{\nu}=S_0\left(\frac{kT}{kT^*}\right)^{5/2},
\end{equation}
where
\begin{equation}
S_0=(27.69\;{\rm mJy})f_{\nu}h^2d_a^2f_b
        \left(\frac{kT^*}{6.4{\rm keV}}\right), 
\end{equation}
and $d_a$ is the dimensionless part of $D_a(z)$: $D_a(z)=(c/H_0)d_a$.
Replacing the gas temperature in equation (10) by the X-ray
luminosity in term of equation (8), we have   
\begin{equation}
S_{\nu}=S_0
        \left(\frac{kT_0}{kT^*}\right)^{5/2}
	\left(\frac{L_{\rm X}}{L_0}\right)^{5B/2}
	(1+z)^{5C/2}.
\end{equation}
These relations constitute a crucial link between the total 
SZ flux in SZ surveys and the cluster distribution characterized
by MF or XLF.

\section{SZ cluster counts: MF}

In the standard model, 
the PS formalism is adopted to describe the MF of
clusters, which provides a reasonable scenario for 
the evolution of abundance of massive dark halos that grow from random-phase
Gaussian initial fluctuations:
\begin{equation}
dn=-\sqrt{\frac{2}{\pi}} \frac{\bar{\rho}}{M} 
    \frac{\delta_c}{\sigma^2}   \frac{d\sigma}{dM}  
    \exp\left(-\frac{\delta_c^2}{2\sigma^2}\right) dM,
\end{equation}
where $\bar{\rho}$ is the mean cosmic density, $\delta_c$ is the 
linear over-density of perturbations that collapsed and virialized at 
redshift $z$, and $\sigma$ is the linear theory variance of the mass density
fluctuation in sphere of mass $M$. Multiplying this MF by
the comoving volume element and integrating over
masses greater than the limit $M_{\rm min}$
 set by the SZ flux $S_{\nu}$ (equation [9])
yield the differential number of SZ clusters:
\begin{equation}
\frac{dN(>S_{\nu})}{dzd\Omega}=\frac{dV}{dzd\Omega}
	\int_{M_{\rm min}(z,S_{\nu})}^{\infty} \left(\frac{dn}{dM}\right) dM,
\end{equation}
where the comoving volume element is
\begin{equation}
\frac{dV}{d\Omega}= \left(\frac{c}{H_0}\right)^3
	\frac{(1+z)^2}{E(z)}d_a^2 dz. 
\end{equation}
The total SZ cluster counts can be obtained by integrating
the above expression over entire redshift space
\begin{equation}
\frac{dN(>S_{\nu})}{d\Omega}=\int_0^{\infty}
	\left(\frac{dV}{dzd\Omega}\right) dz
	\int_{M_{\rm min}(z,S_{\nu})}^{\infty} \left(\frac{dn}{dM}\right) dM.
\end{equation}

Following Viana \& Liddle (1999), we adopt the linear over-density
threshold $\delta_c=1.688$ and the approximate expression for 
the variance of the fluctuation spectrum
filtered on the comoving scale $R=(3M/4\pi\Omega_{\rm M}\rho_c)^{1/3}$  
\begin{equation}
\sigma=\sigma_8(z)\left(\frac{R}{8h^{-1}\;\rm Mpc}\right)^{-\gamma(R)},
\end{equation}
where 
\begin{equation}
\gamma(R)=(0.3\Gamma+0.2)
\left[2.92+\log\left(\frac{R}{8h^{-1}\;\rm Mpc}\right)\right],
\end{equation}
and  $\Gamma$ is the so-called shape parameter.
The quantity $\sigma_8(z)$ is defined by
\begin{equation}
\sigma_8(z)=\frac{\sigma_8}{1+z}
         \frac{g[\Omega_{\rm M}(z),\Omega_{\Lambda}(z)]}
              {g[\Omega_{\rm M},\Omega_{\Lambda}]},
\end{equation}
in which the linear growth factor can be reasonably approximated by
(Carroll, Press \& Turner 1992)
\begin{equation}
g(\Omega_{\rm M},\Omega_{\Lambda})=\frac{5}{2}
	\frac{\Omega_{\rm M}}
 {[\Omega_{\rm M}^{4/7}-\Omega_{\Lambda}+
 (1+\Omega_{\rm M}/2)(1+\Omega_{\Lambda}/70)]}.
\end{equation}
Numerous studies have been devoted to the calibration of 
the present value $\sigma_8$ using primarily CMB anisotropy measurements
and cluster abundances. We adopt the result deduced 
by Eke, Cole \& Frenk (1996) based on the cluster temperature function 
\begin{equation}
\sigma_8=\left\{
\begin{array}{ll}
0.52\Omega_{\rm M}^{-0.52+0.13\Omega_{\rm M}}, & 
                    \Omega_{\rm M}+\Omega_{\Lambda}=1;\\
0.52\Omega_{\rm M}^{-0.46+0.10\Omega_{\rm M}}, & 
                    {\rm open}\;\;\Omega_{\Lambda}=0.
\end{array} \right.
\end{equation}
Another parameter is the overdensity of dark matter with respect to 
the critical density $\rho_c$,  $\Delta_c$, for which we take
the approximate expression fitted by Bryan \& Norman (1998)
\begin{equation}
\Delta_c=\left\{
\begin{array}{ll}
18\pi^2+82x-39x^2, &  \Omega_{\rm K}=0;\\
18\pi^2+60x-32x^2, &  \Omega_{\Lambda}=0,
\end{array} \right.
\end{equation}
where $x=\Omega_{\rm M}(z)-1$. This formula is accurate to within $1\%$ for
$0.1<\Omega_{\rm M}(z)<1$.

\section{SZ cluster counts: XLF}

The reliability of the theoretical prediction for SZ cluster surveys  
from the cluster XLF depends on our current knowledge of 
the dynamical properties of intracluster gas
and the cosmic evolution of the cluster XLF.
Over the past years, various surveys with ROSAT using different X-ray
selection methods have yielded an unprecedented sampling of the 
XLF of local and distant clusters. Essentially, the local XLF and
the XLF at high redshifts out to $z\approx0.8$ are found to be in 
excellent agreement with each other, indicating that bulk of cluster
population have experienced no significant evolution since $z\approx1$
(e.g. Ebeling et al. 1997; Collins et al. 1997; Rosati et al. 1998;
Burke et al. 1997; De Grandi et al. 1999). However, this does not
exclude the possibility that the most luminous clusters 
(typically $L_{\rm X}[0.5$ $- 2$ keV$]>5\times10^{44}$ erg s$^{-1}$)
may become rarer with increasing redshift, at least at $z>0.5$ 
(Rosati et al. 2000).
So, we would like to adopt both non-evolving and evolving XLFs of 
clusters in our theoretical predictions of SZ cluster surveys and
compare their differences.

The differential XLF of clusters can be modeled by the Schechter function 
(e.g. Ebeling et al. 1997; Rosati et al. 2000; etc.)
\begin{equation}
\frac{dn}{dL_{\rm X}}=A\exp(-L_{\rm X}/L_{\rm X}^*)L_{\rm X}^{-\alpha},
\end{equation}
where the cosmic evolution is characterized by 
$A=A_0(1+z)^{\bar{A}}$ and $L_{\rm X}^*=L_{\rm X0}^*(1+z)^{\bar{B}}$.
(Rosati et al. 2000). For the non-evolving XLF, we adopt 
the local XLF constructed by Ebeling et al. (1997) in 
the 0.5 - 2 keV band: 
$A_0=3.32_{-0.36}^{+0.33}\times10^{-7}$ Mpc$^{-3}$ 
($10^{44}$ ergs s$^{-1}$)$^{\alpha-1}$,
$L_{\rm X0}^*=5.70^{+1.29}_{-0.93}\times10^{44}$ ergs s$^{-1}$ and 
$\alpha=1.85^{+0.09}_{-0.09}$, where the Hubble constant is $h=0.5$, and
the errors are $68\%$ confidence limits. For the evolving XLF, we adopt 
the evolution parameters given by Rosati et al. (2000)
for an Einstein-de-Sitter universe: 
$\bar{A}\approx0$ and $\bar{B}=-3$. 
In order to convert the above XLF given in a flat cosmological model
of $\Omega_{\rm M}=1$ and $\Omega_{\Lambda}=0$ into 
the ones in arbitrary cosmological models, we 
demand that the observed number ($dN_{\rm obs}$) of clusters 
in a given redshift interval ($z,z+dz$) be conserved, namely,
\begin{equation}
dN_{\rm obs}  =\left(\frac{dV}{dz}\right)
          dn(L_{\rm X})dz
             =\left(\frac{dV_0}{dz}\right)dn_0(L_{{\rm x},0})dz,
\end{equation}
in which the quantities $dV$, $dn$ and $L_{\rm X}$ 
with and without the subscript
`0' correspond to the Einstein-de-Sitter universe ($\Omega_{\rm M}=1$ and 
$\Omega_{\Lambda}=0$) and other cosmological models, respectively.
This enables us to express the cluster XLF in arbitrary cosmological models
in terms of the XLF in the Einstein-de-Sitter universe:
\begin{equation}
\left(\frac{dn}{dL_{\rm X}}\right)dL_{\rm X}=\left(\frac{dV_0}{dV}\right)
                    \left(\frac{dn_0}{dL_{\rm X,0}}\right) dL_{\rm X,0}.
\end{equation}
Here X-ray luminosities in the two cosmological models are connected by
\begin{equation}
L_{\rm X}=\left[\frac{D_L}{D_{L,0}}\right]^2 L_{\rm X,0},
\end{equation}
where $D_L$ and $D_{L,0}$ are the corresponding luminosity distances.
For nearby clusters, the luminosity distance $D_L$ and the comoving volume 
element $dV$ are insensitive to the cosmological parameters 
$\Omega_{\rm M}$ and $\Omega_{\Lambda}$. So, the local XLF of
the Einstein-de-Sitter universe remains roughly unchanged when scaled to 
arbitrary cosmological models.

Replacing the MF in equations (14) and (16)  by
the cluster XLF and integrating over X-ray luminosity give
the differential and total number of clusters with SZ flux greater than 
$S_{\nu}$:
\begin{equation}
\frac{dN(>S_{\nu})}{dzd\Omega}=\frac{dV}{dzd\Omega}
	\int_{L_{\rm X}(z,S_{\nu})}^{\infty} 
         \left(\frac{dn}{dL_{\rm X}}\right) dLx,
\end{equation}
and
\begin{equation}
\frac{dN(>S_{\nu})}{d\Omega}=
	\int_0^{\infty} \left(\frac{dV}{dzd\Omega}\right) dz
	\int_{L_{\rm X}(z,S_{\nu})}^{\infty} 
         \left(\frac{dn}{dL_{\rm X}}\right) dLx,
\end{equation}
respectively, where the lower limit $L_{\rm X}(z,S_{\nu})$ is determined by
equation (12).

Finally, we come to the observationally determined  
$L_{\rm X}$-$T$ relation for clusters. 
With the improvement of the accuracy in the determination of 
the $L_{\rm X}$-$T$ relations for local and distant clusters over the 
past two decades, it has been well established that the cluster 
$L_{\rm X}$-$T$ relation
exhibits no significant evolution at least out to $z\sim0.8$ 
(e.g. Mushotzky \& Scharf 1997; Della Ceca et al. 2000; Fairly et al. 2000). 
This sets $C\approx0$ in equation (8). 
However, the observationally determined $L_{\rm X}$-$T$ relation
varies with the X-ray luminosity range: The best-fit $L_{\rm X}$-$T$ relation
becomes shallower for X-ray luminous clusters than for fainter ones,
which is probably associated with non-gravitational heating.
Here we would like to work with two X-ray luminosity ranges:
$L_{\rm X}[0.5$ $-$ $2$ keV]
$\geq L_{\rm X}^{\rm h}=3\times10^{44}$ erg s$^{-1}$ and
$L_{\rm X}[0.5$ $-$ $2$ keV]
$\geq L_{\rm X}^{\rm l}=1\times10^{43}$ erg s$^{-1}$.
The former corresponds roughly to the ongoing and upcoming SZ 
cluster surveys (e.g. Grego et al. 2000; Udomprasert et al. 2000),
and the latter may be useful for future SZ poor cluster surveys. 
We determine the corresponding cluster $L_{\rm X}$-$T$ relations
in the 0.5 - 2.0 band  
using the non-exhausted catalog of X-ray clusters compiled by 
Wu et al. (1999) and Xue \& Wu (2000). A total of 86 luminous 
($L_{\rm X}\geq L_{\rm X}^{\rm h}$) and 162 luminous plus faint   
($L_{\rm X}\geq L_{\rm X}^{\rm l}$)
clusters whose measurement uncertainties in both $T$ and
$L_{\rm X}$ are available are used for this analysis. Our best-fit
$L_{\rm X}$-$T$ relations in the 0.5 - 2 keV band are
\begin{eqnarray}
L_{\rm X}=10^{-0.85\pm0.11}T^{1.91\pm0.12},\;\;\;
              L_{\rm X}\geq L_{\rm X}^{\rm h};\\ 
L_{\rm X}=10^{-1.16\pm0.06}T^{2.27\pm0.08},\;\;\;
              L_{\rm X}\geq L_{\rm X}^{\rm l},
\end{eqnarray}
where $L_{\rm X}$ is computed for a flat cosmological model of
$\Omega_{\rm M}=1$ and $h=0.5$, $L_{\rm X}$ and $T$ are in units 
of $10^{44}$ erg s$^{-1}$ and keV, respectively, and the quoted errors
are $68\%$ confidence limits. Note that these relations are apparently 
shallower than the bolometric X-ray luminosity - temperature relation
which reads $L_{\rm bol}=10^{-0.93\pm0.06}T^{2.65\pm0.08}$ for an ensemble of
162 clusters analyzed above. The similar tendency was noticed by 
Markevitch (1998) for the $L_{\rm X}$-$T$ relation 
in the 0.1 - 2.4 band.  In principle, 
the best-fit $L_{\rm X}$-$T$ relations in other 
cosmological models can be correspondingly obtained. Here we'd rather take 
a simple way to estimate the $L_{\rm X}$-$T$ relations for 
other cosmological models in terms of equation (26).

\section{Numerical results and comparison}

Essentially, the PS formalism provides a good approximation of the MF for   
dark halos over entire mass ranges. The theoretically predicted SZ counts 
based on equations (14) and (16) have therefore included the contribution of  
less massive virialized objects satisfying the SZ flux limit $S_{\nu}$ 
(equation [9]). On the other hand, we have set a lower X-ray luminosity limit 
($L_{\rm X}\geq L_{\rm X}^{\rm h}$ or $L_{\rm X}\geq L_{\rm X}^{\rm l}$)
in the estimate of the SZ cluster counts within the framework of the 
cluster XLF.
In order to facilitate our comparison of the SZ cluster counts between
the two different scenarios, the mass thresholds in equations (14) and (16)
should be compatible with the X-ray luminosity thresholds in 
equations (27) and (28).
This can be achieved by combining the $M$-$T$ and $L_{\rm X}$-$T$ relations,
equations (6)-(8). It appears that 
the resultant lower mass limit $M_{\rm min}$ varies with
cosmological model, the cluster $L_{\rm X}$-$T$ relation and 
redshift. For examples, in the case of $\Omega_{\rm M}=1$,
$\Omega_{\Lambda}=0$ and $h=0.5$,  
the typical values of $M_{\rm min}$ at $z=0$ set by 
$L_{\rm X}^{\rm h}=3\times10^{44}$ erg s$^{-1}$ and
$L_{\rm X}^{\rm l}=1\times10^{43}$ erg s$^{-1}$ are
$1.43\times10^{15}M_{\odot}$ and
$1.66\times10^{14}M_{\odot}$,  respectively.

We adopt three cosmological models to perform our numerical 
computations: the standard cold dark matter (SCDM) model,  
a low density, flat CDM model with a nonzero cosmological 
constant of $\Omega_{\Lambda}$ ($\Lambda$CDM) and a low density, 
open CDM model (OCDM), and their parameters are summarized  
in Table 2. In addition, the baryon fraction is assumed to be $f_b=0.1$.
We demonstrate our numerical results using a central frequency of 
$\nu=90$ GHz for the SZ survey, which is roughly the AMIBA range  
(Fan \& Chiueh 2001).
The differential SZ counts of clusters with a flux density  
$S_{\nu}\geq20$ mJy (approximately the AMIBA flux sensitivity) 
and the integrated SZ counts of clusters within redshift $z\leq1.5$ 
are displayed in Figure 1 and Figure 2 for the
three cosmological models and the two X-ray luminosity limits, respectively.
We have extrapolated the cluster XLFs to $z=1.5$, while at lower redshifts the 
SZ counts are primarily governed by the lower X-ray luminosity limit 
$L_{\rm X}^{\rm h}$ or $L_{\rm X}^{\rm l}$. The latter accounts for
the sharp drop of the SZ counts towards $z=0$ especially in the case of
$L_{\rm X}>L_{\rm X}^{\rm l}$, leading the maximum 
location ($z_{\rm max}$) of the differential SZ counts to move to a  
larger redshift as compared with the curves predicted by the PS approach.   
Beyond $z_{\rm max}$ the differential SZ counts given by MF 
follow the well-known dependence on cosmological models
(e.g. Barbosa et al. 1996; Holder et al. 2000):

 \begin{table*}
 \caption{Cosmological models}
 \begin{tabular}{llllll}
 \tableline
 \tableline
 Model & $\Omega_{\rm M}$ & $\Omega_{\Lambda}$ & $h$ & $\Gamma$ & $\sigma_8$ \\
 \tableline
 SCDM & 1    & 0   & 0.5  & 0.5  & 0.52 \\
 $\Lambda$CDM & 0.3  & 0.7 & 0.83 & 0.25 & 0.93 \\
 OCDM & 0.3  & 0   & 0.83 & 0.25 & 0.87 \\
 \tableline
 \end{tabular}
  \end{table*}   

\placefigure{fig1}
\placefigure{fig2}

The SZ cluster counts predicted by the MF, the non-evolving and evolving XLFs
of clusters demonstrate remarkable differences:  Firstly, 
in the scenario of non-evolving XLF much more clusters would be expected 
at high redshifts than those predicted by the cluster MF and 
evolving XLF models.
This explains the significant excess of differential SZ counts 
at high redshifts in the non-evolving XLF model. Secondly, 
as compared with the SZ cluster counts depicted 
by the PS formalism,  the evolving XLF model 
yields an excess of the SZ counts within $z\sim0.7$ but a deficit 
at higher redshifts in $\Lambda$CDM and OCDM models. 
In SCDM model, the opposite situation occurs at high redshifts.
This last point can be understood qualitatively: The dark halos would
experience a rapid evolution in SCDM model than 
in $\Lambda$CDM and OCDM models. 
Alternatively, we cannot exclude the possibility that 
the above deficit of SZ cluster counts at high reshifts  
predicted by the evolving XLF model
in $\Lambda$CDM and OCDM is partially due to our extrapolation of the 
evolving XLF model by Rosati et al. (2000) to high redshifts ($z>1$).
Nonetheless, it is seen that the expectations for SZ cluster counts 
are indeed sensitive to the evolving XLF model. 
Thirdly, the expected SZ cluster counts depend critically on the adopted
$L_{\rm X}$-$T$ relation. The fact that neither of the XLF models can 
precisely reproduce the differential and integrated distributions of the SZ 
clusters predicted by the standard PS formalism in three
popular cosmological models especially in the $\Lambda$CDM arises from 
our shallow $L_{\rm X}$-$T$ relations. Note that the consistency 
in the distributions of SZ cluster counts 
between the MF and XLF models are apparently improved for the 
cluster sample with the lower X-ray luminosity limit 
$L_{\rm X}\geq L_{\rm X}^{\rm l}$ which exhibits a slightly steeper 
$L_{\rm X}$-$T$ relation $L_{\rm X}\propto T^{2.27}$. Recall that 
many previous attempts at reproducing the XLF and number counts of
clusters from the PS formalism 
require a steep bolometric X-ray luminosity - temperature
relation $L_{\rm X}\propto T^{3}$ (e.g. Kitayama \& Suto 1997;
Mathiesen \& Evrard 1998, Borgani et al. 1999). In particular, 
a shallower $L_{\rm X}$-$T$ relation leads to an underestimate of
the local density of X-ray luminous clusters according to
the PS approach (Mathiesen \& Evrard 1998),
which may partially account for the differences in the predicted
SZ clusters between the MF and XLF models at lower redshifts 
shown in Figure 1.  Fourthly, the slope of the integrated 
SZ cluster counts, $\log(dN/d\Omega)/\log S_{\nu}$, depends 
not only on cosmological models as noticed in the literature 
(e.g. Barbosa et al. 1996) but also on cluster XLF.
For X-ray luminous clusters with 
$L_{\rm X}\geq L_{\rm X}^{\rm h}$, the total number of the predicted 
SZ clusters can differ by an order of magnitude  among
the MF, the non-evolving and evolving XLF models. 
In a word, the cosmological applications of SZ cluster counts
may be complicated by the presence of cosmic evolution of the intracluster 
gas which may not been fully accounted for in the cluster MS 
described by the PS formalism.

\section{Discussion and conclusions}

The PS approximation has been widely adopted in the theoretical predictions 
of non-targeted SZ surveys. In principle, 
such a treatment allows one to include the contribution of 
all the virialized halos distributed at different redshifts. 
It appears that the predicted SZ counts  can act as
a sensitive probe of the cosmological parameters such as 
$\Omega_{\rm M}$, $\Omega_{\Lambda}$ and $\sigma_8$
(Barbosa et al. 1996; Holder et al. 2000; 
Springel et al. 2000; Bartlett  2000; Fan \& Chiueh 2001; etc.).
Yet, the expectation for  SZ  surveys within the framework of the 
PS formalism is based on an oversimplified scenario in which 
the isothermal gas is assumed to be in hydrostatic equilibrium with 
and follow the cosmic evolution of the underlying gravitational 
potentials of virialized dark halos. Moreover, 
it does not account for the hot 
diffuse gas associated with large-scale structures or 
distributed beyond the virial radii, or non-gravitational heating effect.
Other work is thus needed to test and implement this purely theoretical 
treatment in order for future non-targeted SZ surveys to become an 
effective probe of the universe.
While a sophisticated treatment of the issue should eventually 
rely on numerical simulations, an approximate yet useful estimate of 
the SZ counts of clusters would be possible by working 
with the observed XLF instead of the theoretically motivated MF.

Current X-ray observations suggest that bulk of 
rich clusters have experienced no significant evolution out to 
$z\approx1$, whereas the most luminous, high redshift clusters are
apparently rarer. With the employment of both non-evolving and evolving 
XLFs of clusters,   together with the observationally determined
$L_{\rm X}$-$T$ relation and the virial theorem which set up a link between
the SZ flux and the cluster XLF,  we have calculated the expected 
SZ cluster counts for two X-ray luminosity limits,
$L_{\rm X}^{\rm h}=3\times10^{44}$ erg s$^{-1}$ and
$L_{\rm X}^{\rm l}=1\times10^{43}$ erg s$^{-1}$ in the 0.5 - 2.0 band.
The non-evolving XLF has resulted in a significant excess of SZ cluster 
counts at high redshift as compared with the evolving XLF.  
Moreover, a slightly steeper $L_{\rm X}$-$T$ relation than the observed one 
($L_{\rm X}\propto T^{2.65}$ in the bolometric band) may be needed to
reproduce the distributions of SZ clusters predicted 
by the standard PS formalism even in the prevailing $\Lambda$CDM model.

A comparison of the predicted SZ number counts of clusters between MF and 
XLF is potentially important for the cosmological application of 
future SZ surveys of clusters, 
e.g. the determination of the cosmological parameters 
($\Omega_{\rm M}$, $\Omega_{\Lambda}$, $\sigma_8$, etc.). 
The discrepancy between the SZ cluster counts predicted by the two 
scenarios shown in the present paper seems to suggest that  
large uncertainties may arise if future SZ cluster counts are directly 
fitted to the theoretical prediction by the standard PS model,
although the present analysis may suffer from other problems 
such as the uncertainties in the cluster XLF and 
$L_{\rm X}$-$T$ relation at high redshift. 
We emphasize that SZ cluster surveys
are uniquely governed by the content of intracluster gas 
and its dynamical and cosmic evolution, which are only indirectly
related with the virialized dark halos. In this regard,  
the cluster XLF provides a better estimate of SZ cluster counts.
So, regardless of the uncertainties in the adopted XLF and
$L_{\rm X}$-$T$ relation
our present expectations for  SZ cluster counts from cluster
XLF can be considered to be a mock SZ observation of clusters
With the improvement of the accuracy in the determinations of 
the cluster XLF and $L_{\rm X}$-$T$ relation, one will be soon able to
have a more precise estimate of the SZ cluster counts and compare it with 
the predictions of the PS approach. It is likely that 
further work should be done before SZ cluster surveys
are used for cosmological purpose.

\acknowledgments
We thank an anonymous referee for constructive comments that 
improved  the presentation of this work. This work was supported by 
the National Science Foundation of China, under Grant 19725311, and
the Ministry of Science and Technology of China, under Grant NKBRSF 
G19990754.



\figcaption{Expectation for SZ counts of rich clusters with   
$L_{\rm X}\geq3\times10^{44}$ erg s$^{-1}$ in the 0.5 - 2.0 band
for SCDM, $\Lambda$CDM and OCDM models.  Upper panels show
the differential counts with SZ flux $S_{\nu}\geq 20$ mJy at $\nu=90$ GHz,
and lower panels are the integrated counts within $z=1.5$. 
Solid lines represent the results predicted by PS formalism, and
dotted and dashed lines are from non-evolving and evolving 
cluster XLFs, respectively.
\label{fig1}}

\figcaption{The same as Figure 1 but with a lower X-ray luminosity limit  
$L_{\rm X}\geq1\times10^{43}$ erg s$^{-1}$ in the 0.5 - 2.0 band.
\label{fig2}}


\begin{references}

\reference{}Barbosa, D., Bartlett, J. G., Blanchard, A., \& Oukbir. J. 
		1996, \aap, 314, 13
\reference{}Bartlett, J. G. 2000, \aap, in press (astro-ph/0001267)
\reference{}Bartlett, J. G., \& Silk, J. 1994, \apj, 423, 12
\reference{}Borgani, S., Rosati, P., Tozzi, P., \& Norman, C. 1999, 
		\apj, 517, 40
\reference{}Bryan, G. L., \& Norman, M. L. 1998, \apj, 495, 80  
\reference{}Burke, D. J., et al. 1997, \apj, 488, L83
\reference{}Carroll, S. M., Press, W. H., \& Turner, E. L. 
		1992, \araa, 30, 499
\reference{}Cavaliere, A., Menci, N., \& Tozzi, P. 1998, \apj, 501, 493
\reference{}Collins, C. A., Burke, D. J., Romer, A. K., Sharples, R. M.,
		\& Nichol, R. C. 1997, \apj, 479, L117
\reference{}De Grandi, S., et al. 1999, \apj, 514, 148 
\reference{}Della Ceca, R., et al. 2000, \aap, 353, 498
\reference{}Ebeling, H., et al. 1997, \apj, 479, L101
\reference{}Eke, V. R., Cole, S., \& Frenk, C. S. 1996, \mnras, 282, 263
\reference{}Evrard, A. E., Metzler, C. A., \& Navarro, J. F. 1996, \apj,
                469, 494     
\reference{}Fan, Z., \& Chiueh, T. 2001, \apj, in press
\reference{}Fairly, B. W., et al. 2000, \mnras, submitted (astro-ph/0003324)
\reference{}Fukugita, M., Hogan, C. J., \& Peebles, P. J. E. 1998,
		\apj, 503, 518
\reference{}Holder, G. P., Mohr, J. J., Carlstrom, J. E., Evrard, A. E., \&
		Leitch, E., M. 2000, \apj, 544, 629
\reference{}Horner, J. D., Mushotzky, R. F., \& Scharf, C. A. 1999, \apj,
                520, 78    
\reference{}Kaiser, N. 1986, \mnras, 222, 323   
\reference{}Kitayama, T., \& Suto, Y. 1997, \apj, 490, 557
\reference{}Kneissl, R. 2000, in Large Scale Structure in the X-ray Universe, 
		ed. M. Plionis and I. Georgantopoulos 
		(Paris: Atlantisciences), 185
\reference{}Markevitch, M. 1998, \apj, 504, 27
\reference{}Mathiesen, B., \& Evrard, A. E. 1998, \mnras, 295, 769
\reference{}Mushotzky, R. F., \& Scharf, C. A. 1997, \apj, 482, L13
\reference{}Ponman, T. J., Cannon, D. B., \&  Navarro, J. F. 1999,
                \nat, 397, 135      
\reference{}Press, W. H., \& Schechter, P. 1974, \apj, 187, 425
\reference{}Rosati, P., Borgani, S., Della Ceca, R., Stanford, A.,
	  	Eisenhardt, P., \& Lidman, C. 2000,
		in Large Scale Structure in the X-ray Universe, 
		ed. M. Plionis and I. Georgantopoulos 
		(Paris: Atlantisciences), 13 
\reference{}Rosati, P., Della Ceca, R., Norman, C., \& Giacconi, R.
		1998, \apj, 492, L21
\reference{}Grego, L., et al. 2000, \apj, submitted (astro-ph/0012067)
\reference{}Seljak, U., Burwell, J., \& Pen, U.-L. 2000, \prd, 
  		submitted (astro-ph/0001120)
\reference{}Springel, V., White, M., \& Hernquist, L. 2000, \apj, 
		submitted (astro-ph/0008133)
\reference{}Udomprasert, P. S., Mason, B. S., \& Readhead, A. C. S. 2000,
		in Constructing the Universe with Clusters of Galaxies,
		ed. F. Durret and D. Gerbal, in press
\reference{}Viana, P. T. P., \& Liddle, A. R. 1999, \mnras, 303, 535
\reference{}Wu, K. K. S., Fabian, A., \& Nulsen, P. E. J. 
		1998, \mnras, 301, L20  
\reference{}Wu, X.-P., Xue, Y.-J., \& Fang, L.-Z. 1999, \apj, 524, 22
\reference{}Xue, Y.-J., \& Wu, X.-P. 2000, \apj, 538, 65
\end{references}
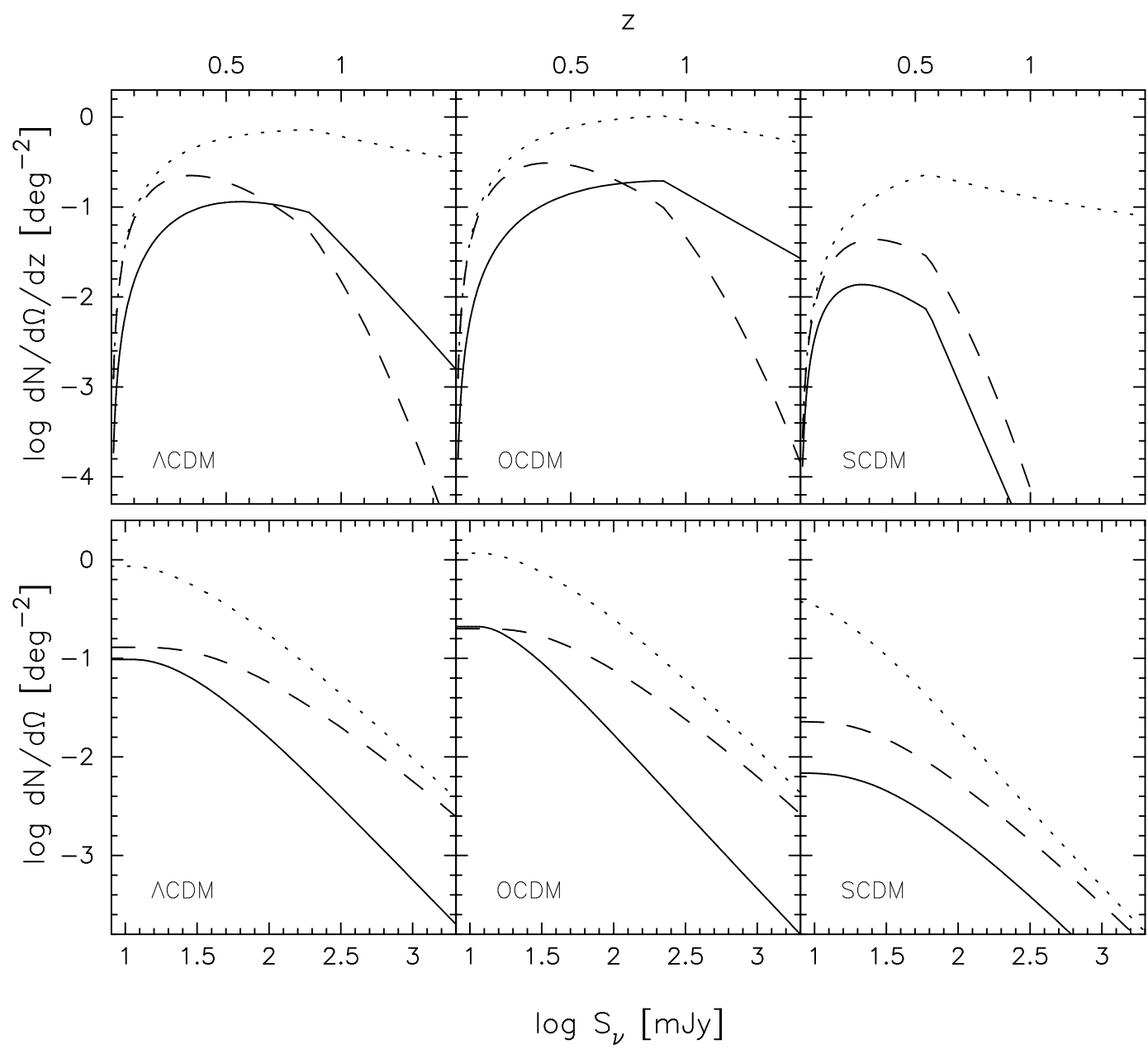
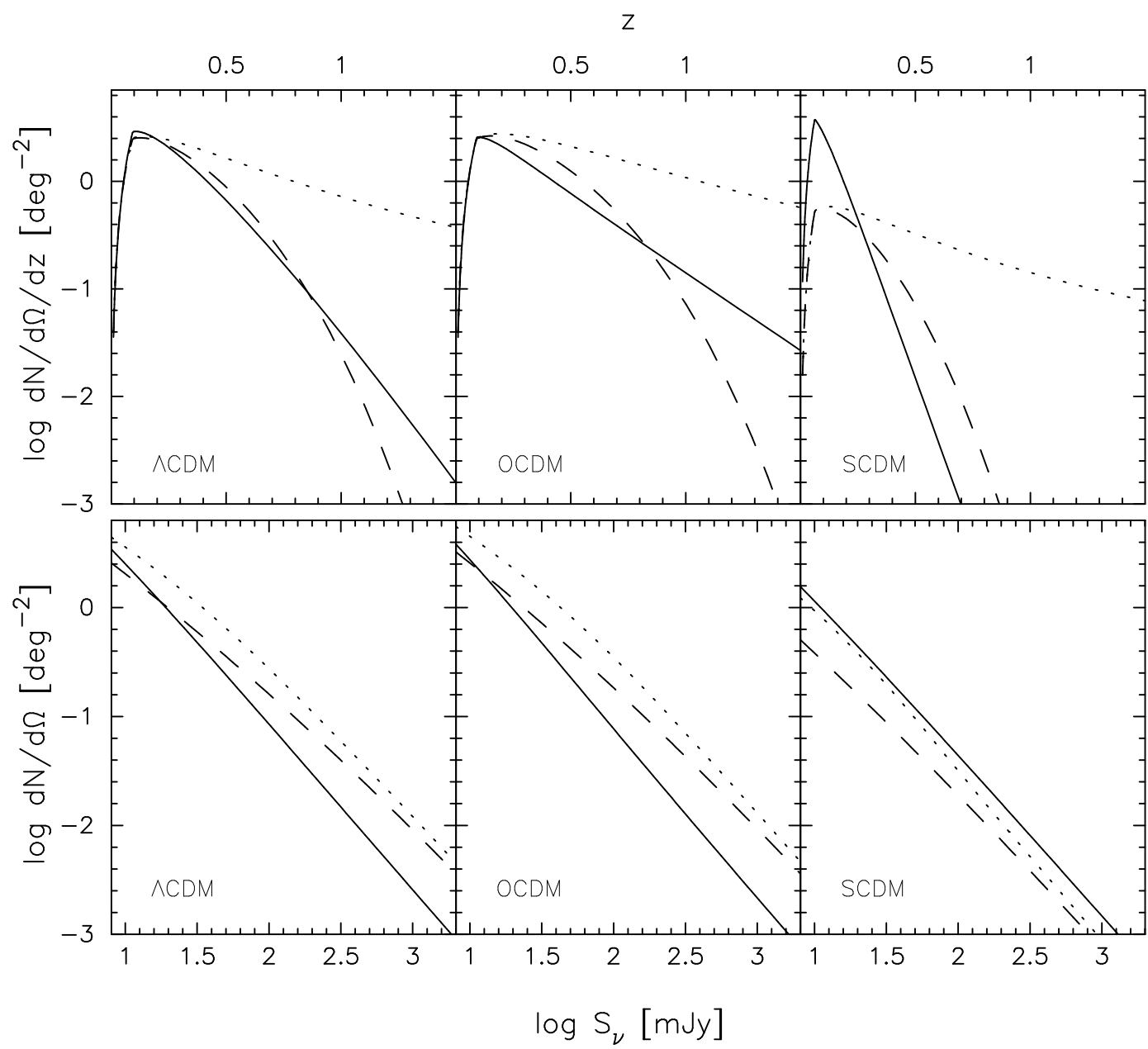
\end{document}